\documentclass[twocolumn,showpacs,prb]{revtex4}
\usepackage{graphicx}
\usepackage{bm}
\usepackage {amsmath}
\usepackage {amssymb}
\usepackage {array}
\usepackage{amsopn} 
\DeclareMathOperator{\trace}{tr}

\newcommand* {\ket}[1]{\ensuremath{\left| {#1} \right\rangle}}
\newcommand* {\matrixelk}[3]{\ensuremath{\langle {#1} | {#2} | {#3} \rangle}}

\newcommand* {\Ds}{\displaystyle}
\newcommand* {\Ts}{\textstyle}
\newcommand* {\vek}[1]{\ensuremath{\bm{#1}}}
\newcommand* {\vekc}[1]{\ensuremath{\bm{\mathcal{#1}}}}
\newcommand* {\kk}{\vek{k}}
\newcommand* {\rr}{\vek{r}}
\newcommand* {\ee}{e}
\newcommand* {\imag}{i}
\newcommand* {\fermi}{\mathrm{F}}
\newcommand* {\bohrmag}{\mu_\mathrm{B}}
\newcommand* {\fj}{\mathfrak{j}}
\newcommand* {\fm}{\mathfrak{m}}

\newcommand{\bk}[4]{\left\langle{\left.{#1\frac{#2}{2}}\right|}
                                {#3\frac{#4}{2}}\right\rangle}
\begin{document}

\title{Spin Density Matrix of Spin-$3/2$ Hole Systems}

\author{R.~Winkler}
\affiliation {Institut f\"{u}r Festk\"orperphysik,
Universit\"{a}t Hannover, Appelstr.~2, D-30167
Hannover, Germany}

\date{February 13, 2004}

\begin{abstract}
  For hole systems with an effective spin $j=3/2$, we present an
  invariant decomposition of the spin density matrix that can be
  interpreted as a multipole expansion.  The charge density
  corresponds to the monopole moment and the spin polarization due
  to a magnetic field corresponds to a dipole moment while heavy
  hole--light hole splitting can be interpreted as a quadrupole
  moment. For quasi two-dimensional hole systems in the presence of
  an in-plane magnetic field $B_\|$ the spin polarization is a
  higher-order effect that is typically much smaller than one even
  if the minority spin subband is completely depopulated. On the
  other hand, the field $B_\|$ can induce a substantial octupole
  moment which is a unique feature of $j=3/2$ hole systems.
\end{abstract}

\maketitle


\section{Introduction}

A powerful approach to the quantum many-body problem is density
functional theory. Hohenberg and Kohn \cite{hoh64} showed that the
(nondegenerate) ground state of a system of interacting particles is
a unique function of the corresponding ground state density $\rho$.
The ground state of the system is thus fully characterized by its
density $\rho$ (Ref.~\onlinecite{dre90}). For electrons with spin $j
= 1/2$ the density $\rho$ must be replaced by a $2\times 2$ spin
density matrix $\vek{\rho}$. It is well-known that $\vek{\rho}$ can
be parameterized by the density $\rho = \trace\vek{\rho}$ and (the
magnitude of) the spin polarization. \cite{dre90} The spin
polarization at a magnetic field $B>0$ is thus an important quantity
for our understanding of many-particle phenomena. Recently, the
interest in this subject has been renewed because experiments have
indicated that the spin polarization of quasi two-dimensional (2D)
systems due to a magnetic field $B_\|$ parallel to the 2D plane
affects the apparent metallic behavior of these systems.
\cite{kra04} Furthermore, the spin polarization is an important
parameter for possible applications in the field of spintronics.
\cite{wol01} In quasi two-dimensional (2D) systems the spin
polarization has been studied by applying a magnetic field $B_\|$
parallel to the 2D plane.
\cite{smi72,oka99,tut02,pud02,zhu03,tut03,pap00a,tut01,pro02,noh03}
If the Zeeman splitting due to $B_\|$
becomes sufficiently large, the minority spin subband is completely
depopulated. In such a situation, electron systems are fully
spin-polarized. This approach has been used by several groups to
study the spin polarization in low-density 2D electron systems.
\cite{tut02,pud02,zhu03,tut03}

Recently, there has been considerable interest in the spin
polarization of quasi 2D hole systems
\cite{pap00a,tut01,pro02,noh03}. Hole systems in the uppermost
valence band of many common semiconductors such as Ge and GaAs are
different from electron systems due to the fact that the hole states
have an effective spin $j=3/2$ (Ref.\ \onlinecite{lut56}). Usually,
hole states in quasi 2D systems are discussed using a basis of
angular momentum $j=3/2$ eigenfunctions with quantization axis
perpendicular to the 2D plane. Subband quantization yields so-called
heavy hole (HH) states with $z$ component of angular momentum $m=
\pm 3/2$ and light hole (LH) states with $m= \pm 1/2$. Often it is
assumed that the spin degree of freedom of these HH and LH states
behaves similar to the spin of $j = 1/2$ electron states. However,
recently it has been shown that the spin polarization of quasi 2D HH
systems in the presence of an in-plane magnetic field behaves rather
different from the more familiar case of $j=1/2$ electron systems.
\cite{win04aa} For example, the spin polarization of quasi 2D HH
systems can change its sign at a finite value of $B_\|$.

Similar to the case of $j=1/2$ electron systems, the fundamental
object for the characterization of $j=3/2$ hole systems is the spin
density matrix $\vek{\rho}$ from which we can derive, e.g., the spin
polarization. The dominant character of the occupied eigenstates of
hole systems depends on the quantization axis of the underlying
$j=3/2$ basis functions. Obviously, all observable quantities such
as the spin polarization may not depend on this choice. Therefore,
it is necessary to formulate the spin density matrix in a way such
that observable quantities can be calculated independent of the
particular choice for the basis functions that are used. It is the
goal of this paper to present such a theory. Using the theory of
invariants \cite{bir74,win03} we will derive an invariant
decomposition of the spin density matrix of $j=3/2$ hole systems
that can be interpreted as a multipole expansion.

\section{Invariant decomposition of the spin density matrix}
\subsection{General theory}

In general, the single-particle states in quasi 2D systems with
effective spin $j$ (the Kohn-Sham states) can be characterized by
multicomponent envelope functions \cite{win03}
\begin{equation}
\label{eq:envelope_fun}
\Psi_{\alpha \sigma \kk_\|} (\rr) =
\frac{\ee^{\imag \kk_\| \cdot \rr_\|}}{2 \pi} \;
\sum_{m} \xi_{\alpha \sigma \kk_\|}^{m} (z) \;
u_{m} (\rr) \; .
\end{equation}
Here $\xi_{\alpha \sigma \kk_\|}^{m} (z)$ denotes the $m$th
component of the envelope function that modulates the band-edge
Bloch function $u_{m} (\rr)$, $m$ labels the $z$ components of $j$,
$\alpha$ is the subband index, $\sigma$ is the spin index, and
$\kk_\| = (k_x, k_y, 0)$ is the in-plane wave vector. The
$(2j+1)$-dimensional Hermitian, positive definite spin density
matrix $\vek{\rho}$ has then the matrix elements (temperature $T=0$)
\begin{equation}
  \label{eq:spin-dens-mat}
  \begin{array}[b]{rl}
    \langle m | m' \rangle =
    & \Ds
    \sum_{\alpha, \sigma}
    \int \! dk_\|^2 \:
    \theta [\pm (E_\fermi - E_{\alpha\sigma} (\kk_\|))]
    \\ & \Ds \times
    \int \! dz  \:
    \xi^{m \,\ast}_{\alpha \sigma \kk_\|} (z) \:
    \xi^{m'}_{\alpha \sigma \kk_\|} (z) \; .
  \end{array}
\end{equation}
Here $E_{\alpha\sigma} (\kk_\|)$ is the spin-split subband
dispersion and $E_\fermi$ is the Fermi energy. The upper (lower)
sign in Eq.\ (\ref{eq:spin-dens-mat}) applies to electron (hole)
systems. We integrate over all orbital degrees of freedom that are
irrelevant in the present discussion. The density is given by the
trace of $\vek{\rho}$. Obviously, the trace is independent of the
particular choice for the set of basis function $u_{m} (\rr)$.

For 3D \emph{homogeneous} hole gases ($B=0$) the matrix $\vek{\rho}$
is proportional to the unit matrix, i.e., we have equal
contributions of HH and LH states because we have no preferential
direction for the quantization axis of angular momentum.
\cite{bob98} (This was overlooked in Ref.\ \onlinecite{end98}.)  For
\emph{inhomogenous} systems such as quasi-2D systems, the matrix
$\vek{\rho}$ has a more complicated structure, even at $B=0$.  In
principle $\vek{\rho}$ contains all information about the ground
state of the system. However, as the Hermitian $4 \times 4$ matrix
$\vek{\rho}$ is characterized by 16 independent parameters, it
appears impractical to characterize a $j=3/2$ system by its density
matrix. Thus, Bobbert \emph{et al.} \cite{bob97} suggested to
parameterize the mixed HH-LH character of inhomogenous hole gases by
the ratio between the LH and HH Fermi wave vectors. Enderlein
\emph{et al.} \cite{end97} proposed to use partial densities for HH
and LH states. The latter approach was criticized by Bobbert
\emph{et al.} \cite{bob98} because the partial hole densities depend
on the set of basis function $u_{m} (\rr)$. Recently, K\"arkk\"ainen
{\em et al.} \cite{kae03} suggested to treat mixed HH-LH systems as
two-component systems where the HH and LH components are essentially
independent of each other. Until today, the interpretation of the $4
\times 4$ spin density matrix $\vek{\rho}$ of particles with $j=3/2$
has remained an open problem.

We show here that the theory of invariants \cite{bir74, win03}
allows to decompose $\vek{\rho}$ for systems with arbitrary point
group $\mathcal{G}$ into symmetry-adapted irreducible tensors which
provide a rigorous yet physically transparent classification of the
different terms in $\vek{\rho}$ --- independent of the particular
choice for the set of basis functions. We assume that the system is
characterized by an $N\times N$ multiband Hamiltonian $\mathcal{H}$.
The basis functions of $\mathcal{H}$ transform according to an
$N$-dimensional representation $\Gamma_\alpha$ of $\mathcal{G}$. For
simplicity, we assume that the representation $\Gamma_\alpha$ is
irreducible. The density matrix $\vek{\rho}$ of this system can be
written in the form
\begin{equation}
  \label{eq:dens_op}
   \vek{\rho} = \sum_{\mu,\nu} M_{\mu\nu} \: \rho_{\mu\nu} \; .
\end{equation}
Here $M_{\mu\nu}$ are $N\times N$ matrices with all elements equal
to zero, except for the element $(\mu,\nu)$ that equals one.  The
components $\{\rho_{\mu\nu}\}$ and the matrices $\{M_{\mu\nu}\}$
transform according to the product representation $\Gamma_\alpha
\times \Gamma_\alpha^\ast$. In general, this representation is
reducible, i.e., we have
\begin{equation}
  \label{eq:red_decomp}
  \Gamma_\alpha \times \Gamma_\alpha^\ast = \sum_\beta \Gamma_\beta \; .
\end{equation}
Equation (\ref{eq:red_decomp}) implies that we can decompose the set
of matrices $\{M_{\mu\nu}\}$ (the components $\{\rho_{\mu\nu}\}$)
into irreducible tensors $\{\vek{M}_\beta\}$
($\{\vek{\rho}_\beta\}$) that transform according to the
representations $\Gamma_\beta$ of $\mathcal{G}$. The
symmetry-adapted expansion of the density matrix $\vek{\rho}$ is
then given by \cite{bir74}
\begin{equation}
  \label{eq:sym_dens_op}
  \vek{\rho}
  = \sum_\beta \vek{M}_\beta \cdot \vek{\rho}_\beta
  \equiv \sum_{\beta,\lambda}
  M_{\beta\lambda} \; \rho_{\beta\lambda}^\ast \; ,
\end{equation}
where $\lambda$ labels the elements of the irreducible
representation $\Gamma_\beta$. The terms $\rho_{\beta\lambda}$ can
be interpreted as the projection of $\vek{\rho}$ on the basis
matrices $M_{\beta\lambda}$. If the matrices $\{ M_{\beta\lambda}
\}$ are orthonormalized we thus have
\begin{equation}
  \label{eq:dens_el}
  \rho_{\beta\lambda} = \trace (M_{\beta\lambda} \,\vek{\rho}) \;.
\end{equation}

According to Eq.\ (\ref{eq:sym_dens_op}), we get a decomposition of
the density matrix $\vek{\rho}$ in terms of invariants
$\vek{M}_\beta \cdot \vek{\rho}_\beta$. Often, these invariants
provide already a clear interpretation of the different terms in
$\vek{\rho}$. A second advantage of the invariant expansion
(\ref{eq:sym_dens_op}) is the following. Using the Clebsch-Gordan
coefficients for the irreducible representations $\{\Gamma_\beta\}$
of $\mathcal{G}$ we can construct \emph{scalar} quantities
$\{\rho_s\}$ from the irreducible tensors $\{\vek{\rho}_\beta \}$
which transform according to the unit representation $\Gamma_1$
(Ref.~\onlinecite{bir74}). These scalars are sums of products of the
elements $\{\rho_{\beta\lambda}\}$. By definition, they are
independent of the particular choice for the set of basis functions
that was used to evaluate the spin density matrix. Therefore, the
scalars $\{\rho_s\}$ represent a convenient set of independent
variables for a parameterization and characterization of the spin
density matrix of a system with point group $\mathcal{G}$.

\subsection{Multipole expansion of the density matrix for systems with
point group $SU(2)$}

For a system with spherical symmetry we have $\mathcal{G} = SU(2)$.
We denote the irreducible representations of $SU(2)$ by
$\mathcal{D}_\fj$ where $\fj$ is the angular momentum. Here the
invariant expansion (\ref{eq:sym_dens_op}) is equivalent to a
multipole expansion of the density matrix. It follows from the
well-known relation for the addition of angular momenta,
\cite{edm60}
\begin{equation}
  \label{eq:add_ang_irep}
  \mathcal{D}_\fj \times \mathcal{D}_{\fj'}^\ast =
  \sum_{\fj'' = |\fj-\fj'|}^{\fj+\fj'} \mathcal{D}_{\fj''} \; ,
\end{equation}
that for a system with angular momentum $j$ the invariant expansion
(\ref{eq:sym_dens_op}) contains multipole moments $\vek{\rho}_\fj$ up
to the order $2j$. The squared magnitude of the $\fj$th multipole
moment is given by \cite{edm60}
\begin{equation}
  \label{eq:mult_mod}
  \rho_\fj^2 = \sum_{\fm = - \fj}^\fj
  (-1)^\fm \: \rho_{\fj\fm} \: \rho_{\fj,-\fm} \; .
\end{equation}
By definition, these scalars are independent of the quantization axis
of angular momentum that is used for the basis functions $u_{m}
(\rr)$ in Eq.\ (\ref{eq:envelope_fun}).

\subsubsection{$j=1/2$ Electron systems}

\begin{table}[tbp]%
  \caption{\label{tab:spn12}Invariant decomposition of the spin density
  matrix of a system with spin $j=1/2$.}
  $\extrarowheight 1ex
   \newcommand{\nl}{\\[0.7ex]}
   \begin{array}{c*{3}{@{\hspace{2em}}l}} \hline \hline
 \mathcal{D}_\fj & \fm & M_{\fj\fm} & \rho_{\fj\fm} \nl \hline

\mathcal{D}_{0}  & 0 & \frac{1}{\sqrt{2}} \, \openone_{2\times 2}
          & \frac{1}{\sqrt{2}}
            \left(\bk{ }{1}{ }{1} + \bk{-}{1}{-}{1}\right) \nl

\mathcal{D}_{1}
          & 1 & -\frac{1}{2}\left(\sigma_x+i\sigma_y\right)
          & -\bk{-}{1}{ }{1} \\[1ex]
          & 0  & \frac{1}{\sqrt{2}} \sigma_z
          & \frac{1}{\sqrt{2}}
            \left(\bk{ }{1}{ }{1} - \bk{-}{1}{-}{1}\right)

  \nl \hline \hline
  \end{array}$
\end{table}

\begin{table}[tbp]%
  \caption{\label{tab:tensop}Tensor operators $\vekc{K}_\fj$ for
  the point group $SU(2)$.}
  $\extrarowheight 1ex
   \newcommand{\nl}{\\[0.7ex]}
   \begin{array}{c*{2}{@{\hspace{0.5em}}l}} \hline \hline
 \mathcal{D}_\fj & \fm & \mathcal{K}_{\fj\fm} \nl \hline

\mathcal{D}_{0}  & 0 & k^2 = \{k_+, k_-\} + k_z^2 \nl

\mathcal{D}_{1}  & 1 & - \frac{1}{\sqrt{2}} B_+ \nl
                 & 0 & B_z \nl

\mathcal{D}_{2}  & 2 & - \frac{1}{2} k_+^2 \nl
                 & 1 & - \{k_+, k_z \} \nl
                 & 0 & \frac{1}{\sqrt{6}} (2 k_z^2 - \{k_+, k_-\})
                 \nl

\mathcal{D}_{3}  & 3 & - \frac{1}{\sqrt{8}} k_+^2 B_+ \nl
                 & 2 & \frac{1}{\sqrt{12}}
                       (k_+^2 B_z + 2\{k_+,k_z\} B_+) \nl
                 & 1 & \frac{1}{\sqrt{120}}
                       (k_+^2 B_- + 2 \{ k_+, k_-\} B_+ - 4 k_z^2 B_+
                        - 8 \{k_+, k_z\} B_z) \nl
                 & 0 & \frac{1}{\sqrt{10}}
                       [(2 k_z^2 - \{k_+, k_-\}) B_z
                        - \{k_+,k_z\} B_- - \{k_-,k_z\} B_+]

  \nl \hline \hline
  \end{array}$
\end{table}

First we apply the above formalism to the well-known case of a
density matrix for systems with angular momentum $j=1/2$. Using
standard angular momentum algebra \cite{edm60} we obtain the
invariant decomposition listed in Table~\ref{tab:spn12}, where
$\sigma_x$, $\sigma_y$, and $\sigma_z$ are the Pauli spin matrices.
We have omitted the terms with $\fm<0$ which can be obtained using
$M_{\fj,-\fm} = (-1)^{\fj+\fm} M_{\fj\fm}^\dagger$ and
$\rho_{\fj,-\fm} = (-1)^{\fj+\fm} \rho_{\fj\fm}^\ast$
(Ref.~\onlinecite{edm60}). We note that the terms $\rho_{\fj\fm}$
can be obtained either from Eq.\ (\ref{eq:dens_el}) or by using the
Clebsch-Gordan coefficients of $SU(2)$ (Ref.~\onlinecite{edm60}). By
summing over the invariants $\vek{M}_\fj \cdot \vek{\rho}_\fj$ it is
easy to check that these invariants provide a decomposition of the
spin density matrix $\vek{\rho}$.

Using Eq.\ (\ref{eq:mult_mod}) we obtain the scalars
\begin{subequations}
  \label{eq:mom_e}
  \begin{eqnarray}
    \rho_0 & = & \textstyle
    \frac{1}{\sqrt{2}} \left(
      \bk{ }{1}{ }{1} + \bk{-}{1}{-}{1} \right) \\[0.5ex]
    \rho_1^2 & = & \textstyle
    \frac{1}{2} \left(\bk{ }{1}{ }{1} - \bk{-}{1}{-}{1}\right)^2
    + 2 \left| \bk{ }{1}{-}{1} \right|^2 \, .
    \label{eq:e_dip}
  \end{eqnarray}
\end{subequations}
Apart from a prefactor $1/\sqrt{2}$, the magnitude of the monopole
moment, $\rho_0$, is equal to the density. (We remark that by
definition $\vek{\rho}_0$ is already a positive scalar that
transforms according to $\mathcal{D}_0$ so that we always have
$\vek{\rho}_0 = \sqrt{\rho_0^2} \equiv \rho_0$.) Apart from a
prefactor $1/2$, the (squared) magnitude of the dipole moment,
$\rho_1^2$, is equal to the (squared) magnitude of the polarization.
Due to the rotational invariance implied by $\mathcal{G} = SU(2)$,
the density matrix is completely characterized by $\rho_0$ and
$\rho_1^2$.

Consistent with the invariant expansion of the density matrix one
can perform an invariant expansion of the Hamiltonian
$\mathcal{H}_{2\times 2}$ of $j=1/2$ electron systems. In lowest
order of $\kk$ and $\vek{B} = (\imag\hbar/e) \kk \times \kk$ we
obtain the tensor operators listed in Table~\ref{tab:tensop}. Here
$k_\pm \equiv k_x \pm \imag k_y$, and $\{\ldots\}$ denotes the
symmetrized product of its arguments, e.g., $\{A,B\} =
\frac{1}{2}(AB+BA)$. We then obtain
\begin{equation}
  \label{eq:ham_e}
  \mathcal{H}_{2\times 2} =
   \frac{\hbar^2 }{\sqrt{2} m^\ast} \vek{M}_0 \cdot \vekc{K}_0
  + \frac{g^\ast}{\sqrt{2}} \bohrmag \vek{M}_1 \cdot \vekc{K}_1
  + V(\rr) \openone_{2\times 2} \:.
\end{equation}
Here $V(\rr)$ is the external potential. The kinetic energy operator
with effective mass $m^\ast$ transforms like a monopole and the
Zeeman term with effective $g$ factor $g^\ast$ transforms like a
dipole. The symbol $\bohrmag$ denotes the Bohr magneton. We see here
that for $j=1/2$ systems the $\fj$th invariant in $\mathcal{H}_{2
\times 2}$ characterizes the dynamics of the $\fj$th invariant of
the spin density matrix. However, we want to emphasize that this
simple correspondence between the different terms in the Hamiltonian
(\ref{eq:ham_e}) and the multipole moments (\ref{eq:mom_e}) is valid
only because we can factorize the eigenfunctions of
$\mathcal{H}_{2\times 2}$ into an orbital part and a spin part. In
general, no such factorization can be performed for $j=3/2$ hole
states. This is the reason why we will find below that hole systems
can have an octupole moment even though the corresponding
Hamiltonian does not contain any octupole term.

\subsubsection{$j=3/2$ Hole systems}

We now perform the invariant decomposition of the spin density
matrix for a system with $j=3/2$. According to Eq.\
(\ref{eq:add_ang_irep}) it contains multipole terms up to $\fj=3$.
Using standard angular momentum algebra \cite{edm60} we obtain the
invariant decomposition listed in Table~\ref{tab:spn32}.
Here $J_x$, $J_y$, and $J_z$ are the angular momentum matrices for
$j=3/2$ and $J_\pm \equiv J_x \pm \imag J_y$. Once again, the
(squared) magnitudes of the multipole moments can readily be
calculated by means of Eq.\ (\ref{eq:mult_mod}). We obtain
\begin{widetext}
\begin{subequations}
  \label{eq:multpol_32}
\begin{eqnarray}
 \rho_0 & = & \Ts
 \frac{1}{2}\left(\bk{ }{3}{ }{3} + \bk{ }{1}{ }{1}
                    + \bk{-}{1}{-}{1} + \bk{-}{3}{-}{3}\right)
 \\[1ex]
 \rho_1^2 & = & \Ts
 \frac{1}{20}\left(3 \bk{ }{3}{ }{3} + \bk{ }{1}{ }{1}
                    - \bk{-}{1}{-}{1} - 3 \bk{-}{3}{-}{3}\right)^2
 + \frac{1}{5} \left| \sqrt{3} \bk{ }{3}{ }{1} + 2 \bk{ }{1}{-}{1}
                    + \sqrt{3} \bk{-}{1}{-}{3} \right|^2
 \\[1ex]
 \rho_2^2 & = & \Ts
 \frac{1}{4}\left(\bk{ }{3}{ }{3} - \bk{ }{1}{ }{1}
                    - \bk{-}{1}{-}{1} + \bk{-}{3}{-}{3}\right)^2
 + \left|\bk{ }{3}{ }{1} - \bk{-}{1}{-}{3} \right|^2
 + \left|\bk{ }{3}{-}{1} + \bk{ }{1}{-}{3} \right|^2
 \\[1ex]
 \rho_3^2 & = & \Ts
 \frac{1}{20}\left(\bk{ }{3}{ }{3} - 3 \bk{ }{1}{ }{1}
               + 3 \bk{-}{1}{-}{1} - \bk{-}{3}{-}{3}\right)^2
 + \frac{2}{5} \left| \bk{ }{3}{ }{1} - \sqrt{3} \bk{ }{1}{-}{1}
                   +  \bk{-}{1}{-}{3} \right|^2
 \nonumber \\ & & \Ts
 + \left|\bk{ }{3}{-}{1} - \bk{ }{1}{-}{3} \right|^2
 + 2 \left|\bk{ }{3}{-}{3} \right|^2 .
\end{eqnarray}
\end{subequations}
\end{widetext}
Apart from a prefactor $1/2$, the monopole moment $\rho_0$ is equal
to the density, and apart from a prefactor $9/20$, the squared
dipole moment $\rho_1^2$ is equal to the squared magnitude of the
polarization. \cite{cart_tens} The quadrupole moment $\vek{\rho}_2$
and the octupole moment $\vek{\rho}_3$ have no equivalent in $j=1/2$
systems. The magnitude $\rho_2$ of the quadrupole moment quantifies
the effect of HH-LH splitting. If only the diagonal elements
$\langle m | m \rangle$ of $\vek{\rho}$ are nonzero, $\rho_2$ is
equal to the difference between the partial HH and LH densities used
by Enderlein \emph{et al.} \cite{end97} (apart from a prefactor
$1/2$). For $j=3/2$ systems, a complete characterization of the
density matrix by means of scalar quantities requires to take into
account additional terms beyond those obtained from Eq.\
(\ref{eq:mult_mod}). They characterize, e.g., the interaction
between different multipole moments. It follows from Eq.\
(\ref{eq:add_ang_irep}) that these scalars are of higher order in
$\langle m | m' \rangle$.

\begin{table*}[tb]%
  \caption{\label{tab:spn32}Invariant decomposition of the spin density
  matrix of a system with $j=3/2$.}
  $\extrarowheight 1ex
   \newcommand{\nl}{\\[0.7ex]}
   \begin{array}{c*{3}{@{\hspace{2em}}l}} \hline \hline
 \mathcal{D}_\fj & \fm & M_{\fj\fm} &
 \rho_{\fj\fm} \nl \hline

\mathcal{D}_{0}  & 0 & \frac{1}{2} \openone_{4\times 4}
          & \frac{1}{2}\left(
              \bk{ }{3}{ }{3} + \bk{ }{1}{ }{1}
            + \bk{-}{1}{-}{1} + \bk{-}{3}{-}{3}\right) \nl

\mathcal{D}_{1}
          & 1 & -\frac{1}{\sqrt{10}} J_+
          & -\sqrt{\frac{3}{10}}
            \big(\bk{ }{1}{ }{3} + \frac{2}{\sqrt{3}} \bk{-}{1}{ }{1}
                  + \bk{-}{3}{-}{1} \big) \nl
          & 0  & \frac{1}{\sqrt{5}} J_z
          &   \frac{1}{2\sqrt{5}}
              \left( 3 \bk{ }{3}{ }{3} + \bk{ }{1}{ }{1}
              - \bk{-}{1}{-}{1} - 3 \bk{-}{3}{-}{3}\right) \nl

\mathcal{D}_{2}
          & 2 & \frac{1}{2\sqrt{6}} J_+^2
          & \frac{1}{\sqrt{2}}
            \left(\bk{-}{1}{ }{3} + \bk{-}{3}{ }{1} \right) \nl
          & 1 & - \frac{1}{\sqrt{6}} \{J_z, J_+ \}
          & \frac{1}{\sqrt{2}}
            \left(- \bk{ }{1}{ }{3} + \bk{-}{3}{-}{1} \right) \nl
          & 0 & \frac{1}{6} \left( 2J_z^2 - J_x^2 - J_y^2 \right)
          &   \frac{1}{2} \left(
              \bk{ }{3}{ }{3} - \bk{ }{1}{ }{1}
            - \bk{-}{1}{-}{1} + \bk{-}{3}{-}{3} \right) \nl

\mathcal{D}_{3}
          & 3 & - \frac{1}{6} J_+^3
          & - \bk{-}{3}{ }{3} \nl
          & 2 & \frac{1}{\sqrt{6}} \{J_z, J_+, J_+ \}
          & \frac{1}{\sqrt{2}}
            \left(\bk{-}{1}{ }{3} - \bk{-}{3}{ }{1} \right) \nl
          & 1 & - \frac{1}{2\sqrt{15}} \left(
                4\{J_z, J_z, J_+ \}
               -  \{J_x, J_x, J_+ \}
                - \{J_y, J_y, J_+ \} \right)
          & - \frac{1}{\sqrt{5}}
            \left(\bk{ }{1}{ }{3} - \sqrt{3} \bk{-}{1}{ }{1}
              + \bk{-}{3}{-}{1} \right) \nl
          & 0 & \frac{1}{3\sqrt{5}} \left(
                2J_z^3 - 3 \{J_x, J_x, J_z \}
                       - 3 \{J_y, J_y, J_z \} \right)
          & \frac{1}{2\sqrt{5}} \left(
              \bk{ }{3}{ }{3} - 3 \bk{ }{1}{ }{1}
              + 3 \bk{-}{1}{-}{1} - \bk{-}{3}{-}{3} \right)
            \nl \hline\hline
  \end{array}$
\end{table*}

The dynamics of $j=3/2$ hole systems is characterized by the $4
\times 4$ Luttinger Hamiltonian $\mathcal{H}_{4\times 4}$
(Ref.~\onlinecite{lut56}). Using Tables~\ref{tab:tensop}
and~\ref{tab:spn32} we express $\mathcal{H}_{4\times 4}$ in terms of
spherical tensor operators \cite{lip70,suz74}
\begin{equation}
  \label{eq:lutt_spher}
  \begin{array}[b]{rcl}
  \mathcal{H}_{4\times 4} & = & \Ds
  - \frac{\hbar^2 \gamma_1}{m_0} \vek{M}_0 \cdot \vekc{K}_0
  - 2\sqrt{5} \kappa \bohrmag \vek{M}_1 \cdot \vekc{K}_1
  \\[2ex] & & \Ds
  + \sqrt{6}\frac{\hbar^2}{m_0} \frac{2\gamma_2 + 3 \gamma_3}{5}
    \vek{M}_2 \cdot \vekc{K}_2
  + \zeta \vek{M}_3 \cdot \vekc{K}_3
  \\[2ex] & & \Ds
  + V(\rr) \openone_{4\times 4} \:,
  \end{array} \hspace*{-2em}
\end{equation}
where $\gamma_1$, $\gamma_2$, and $\gamma_3$ are the Luttinger
parameters and $\kappa$ is the isotropic effective $g$ factor. We
neglect here the small terms with cubic symmetry. They are
considered in the numerical calculations discussed below. Once
again, the magnetic field is acting in spin space as a dipole field
(the operator $\vekc{K}_1$). Obviously, the quadrupole and the
octupole term in $\mathcal{H}_{4\times 4}$ have no equivalent in
$j=1/2$ systems. The quadrupole operator $\vekc{K}_2$ is responsible
for the HH-LH splitting. \cite{lip70} We note that in bulk material,
uniaxial or biaxial strain also gives rise to a splitting of HH and
LH states because analogous to $\vekc{K}_2$ the strain tensor
contains a term $\vekc{K}_2'$ that transforms according to
$\mathcal{D}_2$ so that we get another invariant $\propto \vek{M}_2
\cdot \vekc{K}_2'$ (Ref.~\onlinecite{suz74}). In Eq.\
(\ref{eq:lutt_spher}), the symbol $\zeta$ denotes the prefactor of
the octupole term. Starting from the \mbox{$8\times 8$} Kane model
we obtain in fourth order perturbation theory \cite{win03}
\begin{equation}
  \label{eq:oct_pre}
  \zeta = \frac{\sqrt{2}}{3} \: \frac{e}{\hbar} \: \frac{P^4}{E_0^2}
  \left( \frac{1}{E_0} - \frac{1}{\Delta_0} \right),
\end{equation}
where $P$ is Kane's momentum matrix element, $E_0$ is the
fundamental gap, and $\Delta_0$ is the spin-orbit split-off gap. If
we insert typical values for $P$, $E_0$, and $\Delta_0$ it can be
seen that $\zeta$ is rather small so that the octupole term in
$\mathcal{H}_{4\times 4}$ can safely be neglected. However, we show
below that, nonetheless, the spin density matrix of hole systems can
have a large octupole moment.

\section{Spin density matrix of quasi 2D systems}
\subsection{$j=1/2$ Electron systems}

As an example we consider next the spin density matrix of quasi 2D
electron and HH systems in the presence of an in-plane magnetic
field $B_\|$. For $j=1/2$ electron systems at $B_\| \ge 0$ we can
choose the eigenstates independent of $B_\|$
\begin{equation}
  \label{eq:envelope_qw_e}
\ket{\Psi_{\alpha\pm}^e} =
\frac{\ee^{\imag \kk_\| \cdot \rr_\|}}{2 \pi} \;
\xi_{\alpha}^e (z) \; \frac{1}{\sqrt{2}}
\left( \begin{array}{c} 1 \\ \pm 1 \end{array} \right)_z .
\end{equation}
Using Eq.\ (\ref{eq:e_dip}) we then obtain for the dipole moment
normalized with respect to the total 2D density
\begin{equation}
  \label{eq:multpol_12_qw}
  \tilde{\rho}_1^2 = \left\{
    \begin{array}{l@{\hspace{2em}}c}
  \Ds\frac{1}{2}
  \left( \frac{m^\ast g^\ast \bohrmag B_\|}
              {2 \pi \hbar^2 N} \right)^2 & B_\| < B_p \\[2.4ex]
  \Ds\frac{1}{2} & B_\| \ge B_p
    \end{array} \right.
\end{equation}
where we have assumed that only the lowest orbital subband
$\alpha=1$ is occupied, $N$ denotes the total density in the 2D
system, and $B_p = 2\pi \hbar^2 N / (m^\ast g^\ast \bohrmag)$ is the
magnetic field at which the system becomes fully spin-polarized.
\cite{laythick} In Eq.\ (\ref{eq:envelope_qw_e}) the index $z$ at
the spinor indicates that we have chosen the spin quantization axis
in $z$~direction perpendicular to $B_\|$. Of course, the dipole
moment $\rho_1^2$ may not depend on this seemingly inconvenient
choice. It is easy to check that we get the same result if the spin
quantization axis is chosen parallel to~$B_\|$.

\subsection{$j=3/2$ Heavy hole systems}

Next we consider a 2D HH system in the presence of an in-plane
magnetic field $B_\|$. We write the degenerate HH eigenstates at
$B=0$ in the form \cite{win04aa}
\begin{equation}
  \label{eq:envelope_qw}
\ket{\Psi_{\alpha\pm}^\mathrm{HH}} =
\frac{\ee^{\imag \kk_\| \cdot \rr_\|}}{2 \pi} \;
\xi_{\alpha}^\mathrm{HH} (z) \; \frac{1}{\sqrt{2}}
\left(\begin{array}{c}
1 \\ 0 \\ 0 \\ \pm 1
  \end{array}  \right)_z .
\end{equation}
Here we have assumed that the Luttinger Hamiltonian
(\ref{eq:lutt_spher}) is expressed in a basis of $j=3/2$ angular
momentum eigenfunctions in the order $m=3/2$, $1/2$, $-1/2$, and
$-3/2$; and the index $z$ at the spinor indicates that the
quantization axis of these basis functions is perpendicular to the
2D plane. For simplicity we have neglected HH-LH mixing due to a
nonzero in-plane wave vector $\kk_\|$ and we also neglected the
$\kk_\|$ dependence of the envelope functions
$\xi_{\alpha}^\mathrm{HH} (z)$. These aspects are fully taken into
account in the numerical calculations discussed below. A finite
in-plane magnetic field $B_\|$ gives rise to a mixing of HH and LH
states. In first order degenerate perturbation theory we obtain the
following expressions \cite{win04aa} for the quasi 2D HH states in
an in-plane field $B_\|$
\begin{equation}
\label{eq:envelope_qw_b}
\ket{\Psi_{\alpha\pm}^\mathrm{HH}} =
\frac{\ee^{\imag \kk_\| \cdot \rr_\|}}{2 \pi} \;
\xi_{\alpha}^\mathrm{HH} (z) \; \frac{1}{\sqrt{2}}
\left(\begin{array}{c}
1 \\ - \sqrt{3} \mathcal{K} \mp \sqrt{3} \mathcal{G} \\
\mp \sqrt{3} \mathcal{K} - \sqrt{3} \mathcal{G} \\ \pm 1
  \end{array}  \right)_z ,
\end{equation}
where
\begin{subequations}
  \label{eq:fak}
  \begin{eqnarray}
   \mathcal{K} & \equiv &
   \frac{\kappa \bohrmag B_\|}{E^h_\alpha - E^l_\alpha}, \\
   \mathcal{G} & \equiv & \frac{2\gamma_2 + 3 \gamma_3}{5}
   \frac{2 m_0 \matrixelk{\xi_{\alpha}^\mathrm{HH}}{z^2}
                         {\xi_{\alpha}^\mathrm{LH}} }{\hbar^2}
   \frac{(\bohrmag B_\|)^2}{E^h_\alpha - E^l_\alpha} \;.
   \label{eq:fak_g}
 \end{eqnarray}
\end{subequations}
Here $E^h_\alpha$ ($E^l_\alpha$) is the energy of the $\alpha$th HH
(LH) subband at $B_\|=0$, and we have used the gauge $\vek{A} = (0 ,
-z B_\|, 0)$. The expression for $\mathcal{G}$ in
Ref.~\onlinecite{win04aa} differs slightly from our Eq.\
(\ref{eq:fak_g}) because we have used here the spherical
approximation (\ref{eq:lutt_spher}) of $\mathcal{H}_{4\times 4}$. As
discussed in Ref.~\onlinecite{win04aa} we can often treat the HH-LH
mixing characterized by $\mathcal{K}$ and $\mathcal{G}$ as a small
parameter even if $B_\|$ is so large that only the majority spin
subband is occupied. This is a consequence of the frozen angular
momentum of HH states in quasi 2D systems.

Using Eq.\ (\ref{eq:multpol_32}) we then obtain for the multipole
moments normalized with respect to the total 2D density
\begin{subequations}
  \label{eq:multpol_32_qw}
  \begin{eqnarray}
    \label{eq:dipol_32_qw}
    \tilde{\rho}_1^2 & = &
    \frac{9}{5} \left( \mathcal{K} + \mathcal{G} \, \delta N \right)^2 \\[1ex]
    \tilde{\rho}_2^2 & = &
      \frac{1}{4}
    + 3 \mathcal{K}^2  \left[(\delta N)^2 - 1 \right] \\[1ex]
    \tilde{\rho}_3^2 & = &
      \frac{(\delta N)^2}{2}
    - \mathcal{K}^2 \left[3 (\delta N)^2 - \frac{6}{5} \right]
    - \frac{18}{5} \mathcal{G} \mathcal{K} \, \delta N
    \nonumber \\ & &
    \label{eq:octupol_32_qw}
    - \frac{9}{5} \mathcal{G}^2 \, (\delta N)^2 ,
  \end{eqnarray}
\end{subequations}
where $\delta N = (N_+ - N_-) / (N_+ + N_-)$ denotes the normalized
difference between the 2D densities $N_\pm$ in the majority and the
minority spin subband (with $N = N_+ + N_-$), and we have assumed
that only the lowest HH subband $\alpha=1$ is occupied (which is
usually the case for 2D hole systems). We note that
prior to deriving the multipole moments (\ref{eq:multpol_32_qw}) one
must normalize the perturbed wave functions
(\ref{eq:envelope_qw_b}).

HH-LH mixing gives rise to a dipole moment $\tilde{\rho}_1^2$
proportional to the small parameters $\mathcal{K}$ and $\mathcal{G}$
(Ref.~\onlinecite{win04aa}). In lowest (i.e., zeroth) order of the
HH-LH mixing, the quadrupole moment $\tilde{\rho}_2^2$ is equal to
$1/4$, i.e., $\tilde{\rho}_2^2$ is essentially independent of
$B_\|$, which is a consequence of HH-LH splitting. Finally, we also
get a substantial octupole moment even though we assumed $\zeta=0$
in the Luttinger Hamiltonian (\ref{eq:lutt_spher}). If the minority
spin subband is completely depopulated the normalized octupole
moment equals $1/2$ in lowest (i.e., zeroth) order of the HH-LH
mixing. We note that the Zeeman energy splitting of the HH states in
the presence of an in-plane magnetic field $B_\|$ is proportional to
$B_\|^3$ (Ref.~\onlinecite{win03}). Neglecting the nonparabolicity
of the HH energy dispersion we thus have $\delta N \propto
B_\|^3$.

It is interesting to trace back the physical origin of the multipole
moments $\rho_2$ and $\rho_3$.  In spite of the fact that we do not
have a simple correspondence between the multipoles in the Luttinger
Hamiltonian [Eq.\ (\ref{eq:lutt_spher})] and the multipole moments
$\rho_i$ of the spin density matrix [Eq.\ (\ref{eq:multpol_32_qw})],
we get nonzero multipole moments $\rho_2$ and $\rho_3$ only if the
Hamiltonian (\ref{eq:lutt_spher}) contains a quadrupole term
$\propto \vek{M}_2 \cdot \vekc{K}_2$. If the quadrupole term in Eq.\ 
(\ref{eq:lutt_spher}) were zero the HH eigenstates
(\ref{eq:envelope_qw}) at $B=0$ were degenerate with the
corresponding LH states. Similar to the electron states
(\ref{eq:envelope_qw_e}) these four degenerate states could be
factorized into a scalar orbital part times a spinor that depends
only on $\vek{B}$, and we would have $\rho_2 = \rho_3 = 0$. To the
best of our knowledge, the quadrupole term $\propto \vek{M}_2 \cdot
\vekc{K}_2$ is a unique feature of the Hamiltonian for the extended
valence band states in semiconductors. \cite{lut56,lip70,suz74}

It is instructive to check that the scalars $\tilde{\rho}_i^2$ in
Eq.\ (\ref{eq:multpol_32_qw}) are indeed independent of the
quantization axis $\hat{\vek{e}}$ of angular momentum used in Eqs.\
(\ref{eq:envelope_qw}) and (\ref{eq:envelope_qw_b}). If
$\hat{\vek{e}}$ is chosen parallel to $\vek{B}_\|$ Eq.\
(\ref{eq:envelope_qw_b}) must be replaced by \cite{gol95}
\begin{subequations}
\label{eq:envelope_qw_bb}
\begin{eqnarray}
\ket{\Psi_{\alpha +}^\mathrm{HH}} & = &
\frac{\ee^{\imag \kk_\| \cdot \rr_\|}}{2 \pi} \;
\xi_{\alpha}^\mathrm{HH} (z) \; \frac{1}{2}
\left(\begin{array}{c}
-1 + 3 \mathcal{K} + 3 \mathcal{G} \\ 0 \\
\sqrt{3} (1 + \mathcal{K} + \mathcal{G}) \\ 0
  \end{array}  \right)_x \\
\ket{\Psi_{\alpha -}^\mathrm{HH}} & = &
\frac{\ee^{\imag \kk_\| \cdot \rr_\|}}{2 \pi} \;
\xi_{\alpha}^\mathrm{HH} (z) \; \frac{1}{2}
\left(\begin{array}{c}
0 \\ \sqrt{3} (- 1 + \mathcal{K} - \mathcal{G}) \\
0 \\ 1 + 3 \mathcal{K} - 3 \mathcal{G}
  \end{array}  \right)_x .
\end{eqnarray}
\end{subequations}
In this basis the dominant spinor components of the HH states have
$|m|=1/2$. Nonetheless, we obtain the multipole moments
(\ref{eq:multpol_32_qw}).

\subsection{Numerical results}

\begin{figure}[t]
  \includegraphics[height=150mm]{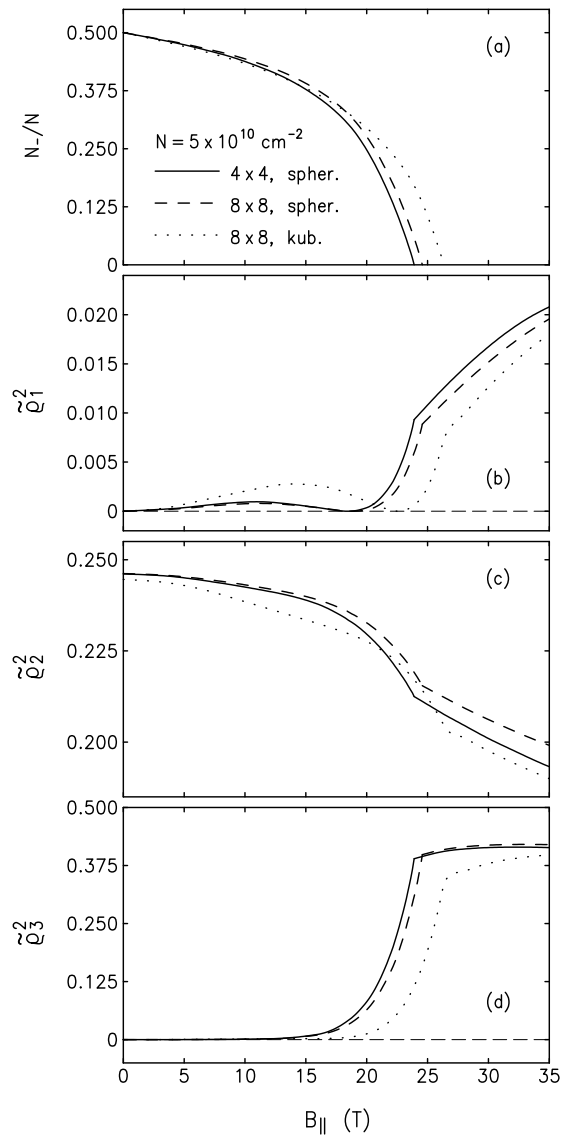}
  \caption{\label{fig:mumo}(a) Normalized spin subband density
  $N_- / N$ of the HH minority spin subband and (b-d) normalized
  multipole moments $\tilde{\rho}_i^2$ as a function of the in-plane
  magnetic field $B_\|$ calculated self-consistently for a symmetric
  (100) GaAs-Al$_{0.3}$Ga$_{0.7}$As quantum well with hole density
  $N = 5 \times 10^{10}$~cm$^{-2}$ and well width $w=150$~{\AA}. The
  solid lines have been obtained by means of the spherical
  approximation (\ref{eq:lutt_spher}) of the Luttinger Hamiltonian.
  The dashed (dotted) lines correspond to the spherical (cubic)
  approximation of the $8 \times 8$ Kane Hamiltonian, respectively.}
\end{figure}

To confirm the qualitative results in Eq.\ (\ref{eq:multpol_32_qw})
we show in Fig.~\ref{fig:mumo} the self-consistently calculated
multipole moments $\tilde{\rho}_i^2$ as a function of the in-plane
magnetic field $B_\|$ for a symmetric (100)
GaAs-Al$_{0.3}$Ga$_{0.7}$As quantum well with hole density $N = 5
\times 10^{10}$~cm$^{-2}$ and well width $w=150$~{\AA}. The
numerical calculations follow Ref.~\onlinecite{win03}.  We evaluate
Eq.\ (\ref{eq:spin-dens-mat}) by means of analytic quadratic
Brillouin zone integration. \cite{win93} The solid lines in
Fig.~\ref{fig:mumo} have been obtained using the spherical
approximation (\ref{eq:lutt_spher}) of the Luttinger Hamiltonian
$\mathcal{H}_{4\times 4}$. However, we have neglected the small
octupole term in Eq.\ (\ref{eq:lutt_spher}). For comparison we also
show the results based on the $8 \times 8$ Kane Hamiltonian
$\mathcal{H}_{8\times 8}$ which includes implicitly the octupole
term. The dashed lines in Fig.~\ref{fig:mumo} are based on the
spherical approximation of $\mathcal{H}_{8\times 8}$ while the
dotted lines also take into account the cubic terms. \cite{win03}

We can see in Fig.~\ref{fig:mumo} that the complete depopulation of
the HH minority spin subband due to $B_\|$ does not imply full spin
polarization of the system. \cite{win04aa} In Fig.~\ref{fig:mumo}
(solid lines), the minority spin subband is completely depopulated
at $B_D \approx 23.9$~T while $\tilde{\rho}_1^2 (B_D) \approx
0.0093$. The latter value is much smaller than $9/20 = 0.45$, the
value of $\tilde{\rho}_1^2$ in a fully spin-polarized HH system. The
zero of $\tilde{\rho}_1^2 (B_\|)$ at $B_\| \approx 18.4$~T reflects
a sign reversal of the spin polarization vector which is a unique
feature of 2D HH systems. \cite{win04aa} The derivative of
$\tilde{\rho}_1^2 (B_\|)$ is discontinuous at $B_\| = B_D$. At $B_\|
= 0$ the quadrupole moment $\tilde{\rho}_2^2$ is slightly smaller
than $1/4$, the value of $\tilde{\rho}_2^2$ in a pure HH system.
This is a consequence of the $\kk_\|$-induced HH-LH mixing which was
fully taken into account in Fig.~\ref{fig:mumo}. For $B_\| > 0$ we
observe only a small decrease of $\tilde{\rho}_2^2$. This is due to
the fact that the HH-LH splitting $E^h_1 - E^l_1 \approx 6.7$~meV
[i.e., the denominators in Eq.\ (\ref{eq:fak})] is the largest
energy scale in the system so that the HH states have a frozen
angular momentum perpendicular to the 2D plane. For comparison, we
note that the Zeeman energy splitting of the HH subband at $B_D$ is
$\sim 0.4$~meV. The spin polarization due to $B_\|$ is thus a
higher-order effect in 2D HH systems.

It is remarkable that the octupole moment $\tilde{\rho}_3^2$ at
$B_\| = B_D$ is close to $1/2$, the largest possible value of
$\tilde{\rho}_3^2$ in a 2D HH system. This value is essentially
independent of whether we use $\mathcal{H}_{4\times 4}$ without an
octupole term or whether we use $\mathcal{H}_{8\times 8}$ which
includes implicitly the octupole term $\propto \vek{M}_3 \cdot
\vekc{K}_3$. For $B_\| \ge B_D$, the octupole moment
$\tilde{\rho}_3^2 (B_\|)$ remains essentially constant, consistent
with Eq.\ (\ref{eq:octupol_32_qw}). These findings suggest that an
in-plane magnetic field $B_\|$ provides an efficient tool to study
2D HH systems with a large octupole moment but with a small dipole
moment (i.e., with a small spin polarization). We can obtain a 2D HH
system with a large spin polarization but with a small octupole
moment if the magnetic field is applied perpendicular to the 2D
plane.

Figure~\ref{fig:mumo} shows that the cubic terms are only small
corrections in $\mathcal{H}_{8\times 8}$. They can safely be
neglected in a discussion of the main features in
Fig.~\ref{fig:mumo}. This is consistent with the fact that the cubic
terms in $\mathcal{H}_{8\times 8}$ (and $\mathcal{H}_{4\times 4}$)
are proportional to $\gamma_3 - \gamma_2$ which is typically a small
parameter. \cite{lip70} On the other hand, it was important for the
results shown in Fig.~\ref{fig:mumo} that the numerical calculations
took into account HH-LH mixing due to a nonzero in-plane wave
vector~$\kk_\|$. As the analytical expressions in Eq.\
(\ref{eq:multpol_32_qw}) neglect this effect, we do not present a
direct comparison between Eq.\ (\ref{eq:multpol_32_qw}) and the
numerical results in Fig.~\ref{fig:mumo}. We remark that it is
straightforward to include the $\kk_\|$-induced HH-LH mixing in Eq.\
(\ref{eq:multpol_32_qw}). However, we do not reproduce here the
lengthy formulas.

\section{Other systems and outlook}

The numerical calculations shown in Fig.~\ref{fig:mumo} have
neglected many-particle effects beyond the Hartree approximation.
These effects will be important for a quantitative analysis of the
spin multipole moments in the density regime discussed here.
However, to the best of our knowledge an appropriate theory for
particles with $j=3/2$ is currently not available. It represents a
formidable task to develop such a theory. We note that the symmetry
arguments developed here are not affected by many-particle effects.
The multipole expansion of the spin density matrix can thus provide
a useful starting point for a more quantitative theory. We expect that
the essential features in Fig.~\ref{fig:mumo} will be confirmed by
such an investigation.

Our results can readily be generalized to spherically symmetric
systems with arbitrary $j$, both half integer and integer,
\cite{edm60} as well as to systems characterized by the
crystallographic point groups. \cite{kos63} As an example, we
consider a $j=1/2$ electron system in a spatial environment with
point group $\mathcal{G} = C_{2v}$. Here the $2 \times 2$ spin
density matrix is completely characterized by five independent
scalars
\begin{subequations}
  \label{eq:dens_c2v}
  \begin{eqnarray}
    \rho_0 & = & \textstyle
      \bk{ }{1}{ }{1} + \bk{-}{1}{-}{1} \\[0.5ex]
    \rho_1 & = & \textstyle
    \left| \bk{ }{1}{-}{1} + \bk{-}{1}{ }{1} \right|^2 \\[0.5ex]
    \rho_2 & = & \textstyle
    \left| \bk{ }{1}{-}{1} - \bk{-}{1}{ }{1} \right|^2 \\[0.5ex]
    \rho_3 & = & \textstyle
    \left| \bk{ }{1}{ }{1} - \bk{-}{1}{-}{1} \right|^2 \\[0.5ex]
    \rho_4 & = & \textstyle
    \left( \bk{ }{1}{ }{1} - \bk{-}{1}{-}{1} \right)
    \left( \bk{ }{1}{-}{1} + \bk{-}{1}{ }{1} \right)
    \nonumber \\ & & \textstyle \times
    \left( \bk{ }{1}{-}{1} - \bk{-}{1}{ }{1} \right) .
  \end{eqnarray}
\end{subequations}
Each term in Eq.\ (\ref{eq:dens_c2v}) has a clear physical
interpretation. Once again, the scalar $\rho_0$ is the charge
density. The scalars $\rho_1$, $\rho_2$, and $\rho_3$ describe the
magnitude of the spin polarization in $x$, $y$, and $z$ direction,
respectively. These quantities must be distinguished from each other
in a system with point group $C_{2v}$. Finally, $\rho_4$ describes
the relative sign of the spin polarization in $x$, $y$, and $z$
direction. The five scalars $\rho_i$ thus form a decomposition of
the four independent parameters of a Hermitian $2\times 2$ matrix
into their magnitudes and their relative signs. For the explicit
formulas in Eq.\ (\ref{eq:dens_c2v}) we have assumed that the
angular momentum quantization axis of the basis functions is
parallel to the symmetry axis of the point group $C_{2v}$. We
emphasize that the magnitude of the scalars $\rho_i$ in Eq.\
(\ref{eq:dens_c2v}) does not depend on this choice.

Finally, we want to briefly address the question of how one can
measure the multipole moments of the spin density matrix of hole
systems. The polarization of $j=1/2$ electron systems can be probed
by measuring the hyperfine interaction between the electrons and the
atomic nuclei. Electrons in the conduction band of common
semiconductors like GaAs originate in $s$-like atomic orbitals.
Therefore, the electrons have a large probability density at the
atomic nuclei so that the hyperfine interaction is large. The
$j=3/2$ hole states in the valence band, on the other hand,
originate in $p$-like atomic orbitals which have a vanishing
probability density at the atomic nuclei. The hyperfine interaction
is therefore much weaker \cite{jac75} so that it will be difficult
to use this technique to measure the polarization of $j=3/2$ hole
states.

Indirect evidence for the unusual features of the spin density
matrix of hole systems can be obtained by varying the confinement of
quasi 2D HH systems perpendicular to the 2D plane. This will change
the HH-LH splitting $E^h_\alpha - E^l_\alpha$ in the denominators of
Eq.\ (\ref{eq:fak}). Here our results are in good qualitative
agreement with the trends observed in recent experiments.
\cite{pap00a,tut01,win03} A detailed comparison with experimental
results will be the subject of a future publication.


\begin{acknowledgments}
  The author appreciates stimulating discussions with A.~H.\
  MacDonald, E.~Tutuc and M.~Shayegan. This work was supported by
  BMBF.
\end{acknowledgments}


\begin{thebibliography}{32}
\bibitem[{car()}]{cart_tens}
\bibinfo{note}{Our definition of the polarization based on spherical tensors is
  equivalent to the approach in Ref.~\onlinecite{win04aa} based on cartesian
  tensors. We remark that one can convert the cartesian components
  $(\mathcal{V}_x, \mathcal{V}_y, \mathcal{V}_z)$ of a vector into spherical
  components $(\mathcal{V}_1, \mathcal{V}_0, \mathcal{V}_{-1} )$ using
  \cite{edm60} \begin{subequations} \begin{eqnarray} \mathcal{V}_{\pm 1} & = &
  \mp \frac{1}{\sqrt{2}} (\mathcal{V}_x \pm \imag \mathcal{V}_y) \\
  \mathcal{V}_0 & = & \mathcal{V}_z \: . \end{eqnarray} \end{subequations}}
\end{thebibliography}
\end{document}